# Quantum Computing Based Design of Multivariate Porous Materials

Shinyoung Kang[1], Younghun Kim[1], Jihan Kim[1]*

[1]Department of Chemical and Biomolecular Engineering, Korea Advanced Institute of Science and Technology, 291 Daehak-ro, Yuseong-gu, Daejeon 34141, Republic of Korea.

*Corresponding author. Email: jihankim@kaist.ac.kr (J. K.)

## Abstract

Multivariate (MTV) porous materials exhibit unique structural complexities based on diverse spatial arrangements of multiple building block combinations. These materials possess potential synergistic functionalities that exceed the sum of their individual components. However, the exponentially increasing design complexity of these materials poses challenges for accurate ground-state configuration prediction and design. To address this, a Hamiltonian model was developed for quantum computing that integrates compositional, structural, and balance constraints, enabling efficient optimization of the MTV configurations. The model employs a graph-based representation to encode linkers as qubits. To validate our model, a variational quantum circuit was constructed and executed using the Sampling VQE algorithm. Simulations on experimentally known MTV porous materials successfully reproduced their ground-state configurations, demonstrating the validity of our model. Furthermore, VQE calculations were performed on real quantum hardware for validation purposes, signaling a first step toward a practical quantum algorithm for the rational design of porous materials.

## Teaser

Utilizing quantum computing with a graph-based Hamiltonian model for the prediction of ground-state configurations in complex multivariate porous materials design

## Introduction

Multivariate (MTV) porous materials contain multiple distinct chemical building units within the same framework. With the ongoing interest in expanding the functionality of porous material (1-7), the incorporation of multiple building blocks within a single framework has led to even larger amount of design freedom and property enhancement compared to pristine porous materials (8-12). For instance, a series of MTV-MOFs known as MUF-7 were developed to demonstrate varied pore distributions and exceptional catalytic capabilities (13). A mixed-linker 2D MOF with copper metal and two trigonal linkers, tetrahydroxy-1,4-quinone (THQ) and 2,3,6,7,10,11-hexahydrotriphenylene (HHTP), was designed to exhibit modulated conductivity and high porosity, both essential qualities for electronic applications such as gas sensing (14). Mixed-linker 2D COFs with ordered pores was proposed to show that orderly and balanced linker arrangements are key to achieving high material stability and functionality (15). Despite these advancements, the number of MTV porous materials remains relatively small due to experimental challenges that arise from the difficulty of obtaining crystal growths and the complexity of incorporating multiple building blocks into one coherent structure (8). With increasing number of metal nodes and linkers, the structural complexity scales exponentially and as such, it becomes impossible to pre-design MTV porous materials for large number of building blocks, which serves as a hindrance to fully explore the search space of these MTV porous material structures.

With this in mind, it is conceivable that computational design can facilitate the search for MTV porous materials by providing blueprints for ground-state configurations. When it comes to *in silico* porous material generation, the top-down approach is commonly employed where given the topology, suitable building blocks are selected to fill in the unit cell (16-18). As such, many research groups have utilized the top-down approach to construct hypothetical structures, and one can imagine using a similar approach to build MTV porous materials with large number of metal nodes and linker types. However, designing such topologically well-ordered linker arrangements using this method becomes increasingly intractable as the problem complexity grows. For example, in hcb topology containing 32 linker sites, the inclusion of eight distinct MTV linkers at some fixed ratio leads to 7.8 quadrillion unique combinatorial structures. This extensive number of potential structures makes it impossible to use any of the existing classical methods to explore the vast search space of MTV porous materials. Therefore, a novel approach is required to traverse through the possible configuration space for the MTV porous materials.

One possible solution that can be used to tackle this issue is through quantum computing. Unlike classical computers, which use bits as their basic unit of computation, quantum computers operate based on the principles of quantum mechanics, utilizing quantum bits (qubits) (19). Qubits possess unique properties, such as superposition and entanglement, enabling quantum algorithms to explore the vast solution space in parallel (20). This capability makes quantum computing particularly well-suited to solve complex NP-hard combinatorial optimization problems (21), which include the well-known traveling salesmen problem that typically requires exponential time to solve using classical brute-force methods (22). Similarly, designing MTV porous materials can be seen as an NP-hard combinatorial optimization problem, given that the number of possible configurations grows exponentially with the increasing number of building blocks and topological sites.

Previously, there have been few studies in the field of chemistry and material sciences that have used quantum computing algorithms to identify the optimal chemical configurations. A quantum optimization algorithm to obtain low-energy conformations of protein models was proposed to search for optimal on-lattice heteropolymers (23). The algorithm incorporates a Hamiltonian that encodes the hydrophobic-polar lattice model, one of the simplest coarse-grained models for protein folding, to search the vast search space of possible conformations (23). To add to the effort for searching optimal heteropolymers, this coarse-grained protein model was extended to a tetrahedral lattice for branched heteropolymers with few monomers by using a two-centered coarse-grained description of amino acids to represent the protein sequence (24). Further exploration was conducted on quantum algorithms in bioinformatics, specifically for mRNA codon optimization (25). This study introduced a more efficient variational quantum eigensolver (VQE)-based encoding method for mRNA codon optimization that halves the qubit requirement, enabling the execution of longer sequences on current quantum processors and producing results closely aligned with exact solutions, thus making the algorithm practical for existing quantum hardware (25). Despite these advancements, to the best of our knowledge, no one has devised a quantum computing algorithm to identify ground-state chemical configurations for porous materials.

In this work for the first time, we propose a Hamiltonian model for quantum computers to design MTV porous materials. By directly embedding compositional, structural, and balance constraints into the Hamiltonian, and representing the topological information of reticular frameworks as a graph-based structure, the proposed quantum algorithm enables efficient exploration of MTV porous material configurations that satisfy all predefined design requirements (Fig. 1). Our model was validated using a variational quantum circuit executed with the quantum algorithm in IBM Qiskit (26). Simulations of experimentally known MTV materials, including Cu-THQ-HHTP, Py-MV-DBA-COF, MUF-7, and SIOC-COF2, successfully reproduced their ground-state configurations, confirming the accuracy of the model. Additionally, the extensibility of this Hamiltonian model was discussed, showcasing its potential for simulating increasingly complex MTV structures as quantum hardware and algorithms continue to advance. This approach utilizes quantum computing's potential to solve NP-hard combinatorial problems, providing a novel framework for optimizing complex MTV porous material architectures beyond the reach of classical methods.

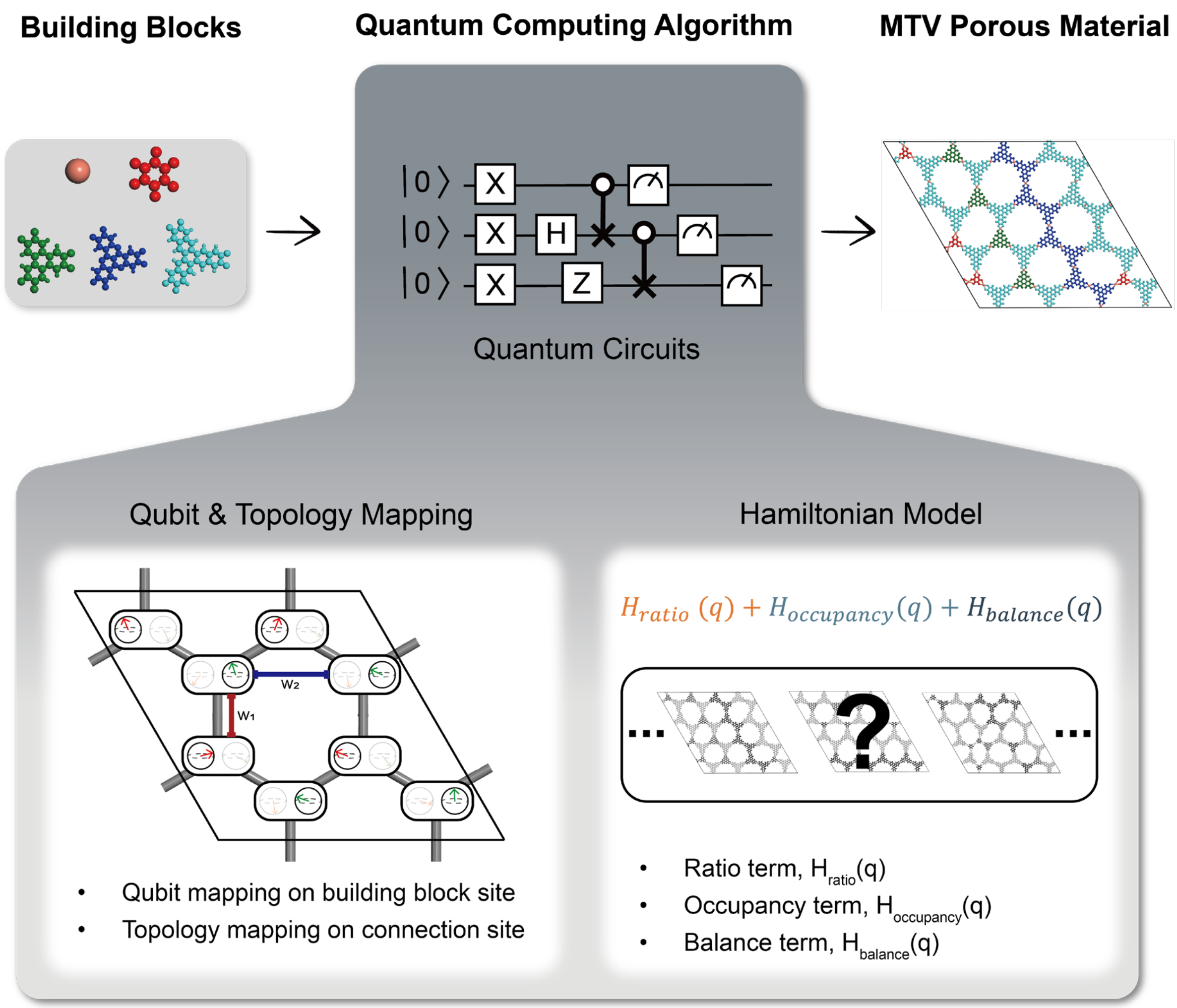

**Fig. 1.** Overall schematics of the quantum computing algorithm to generate feasible MTV porous materials. The algorithm consists of two mapping schemes (qubit mapping and topology mapping) to allocate building blocks in a given connectivity. Different configurations go through a predetermined Hamiltonian, which is comprised of a ratio term, occupancy term, and balance term, to capture the most feasible MTV porous material.

## Results

### The Qubit Representation

To effectively use a quantum computer to navigate through the vast material space of the MTV porous frameworks, the reticular nature of the porous material must be mapped into the qubit representations. In our encoding scheme, the number of qubits, $\mathrm{n}_{qubits}$, is determined by the product of the (1) number of linker types, $|t|$, and the (2) number of linker sites in a defined unit cell, $N_i$, such that $\mathrm{n}_{qubits} = |\mathrm{t}| \times \mathrm{N}_i$. Each qubit represents whether a specific linker type occupies a particular linker site and is labeled as $q_i{}^t$ where the subscript $i$ indicates the linker site, and the superscript $t$ denotes the type of linker.

As a test case, we applied our encoding method to the Cu-THQ-HHTP MOF (14) system. This is a two-dimensional MOF that contains eight linker sites and two linker types (THQ and HHTP) which leads to a total allocation of 16 qubits labeled as $q_0{}^{THQ}, q_0{}^{HHTP}, \ldots, q_7{}^{THQ}, q_7{}^{HHTP}$. A qubit state of 1 (e.g. $q_0{}^{THQ} = 1$) indicates the presence of a THQ linker at site 0, while a state of 0 means that THQ is absent in that site. This encoding allows us to represent every possible configuration of MTV linkers within the unit cell as a unique qubit state. Figure 2A illustrates this qubit representation applied to the defined Cu-THB-HHTP framework.

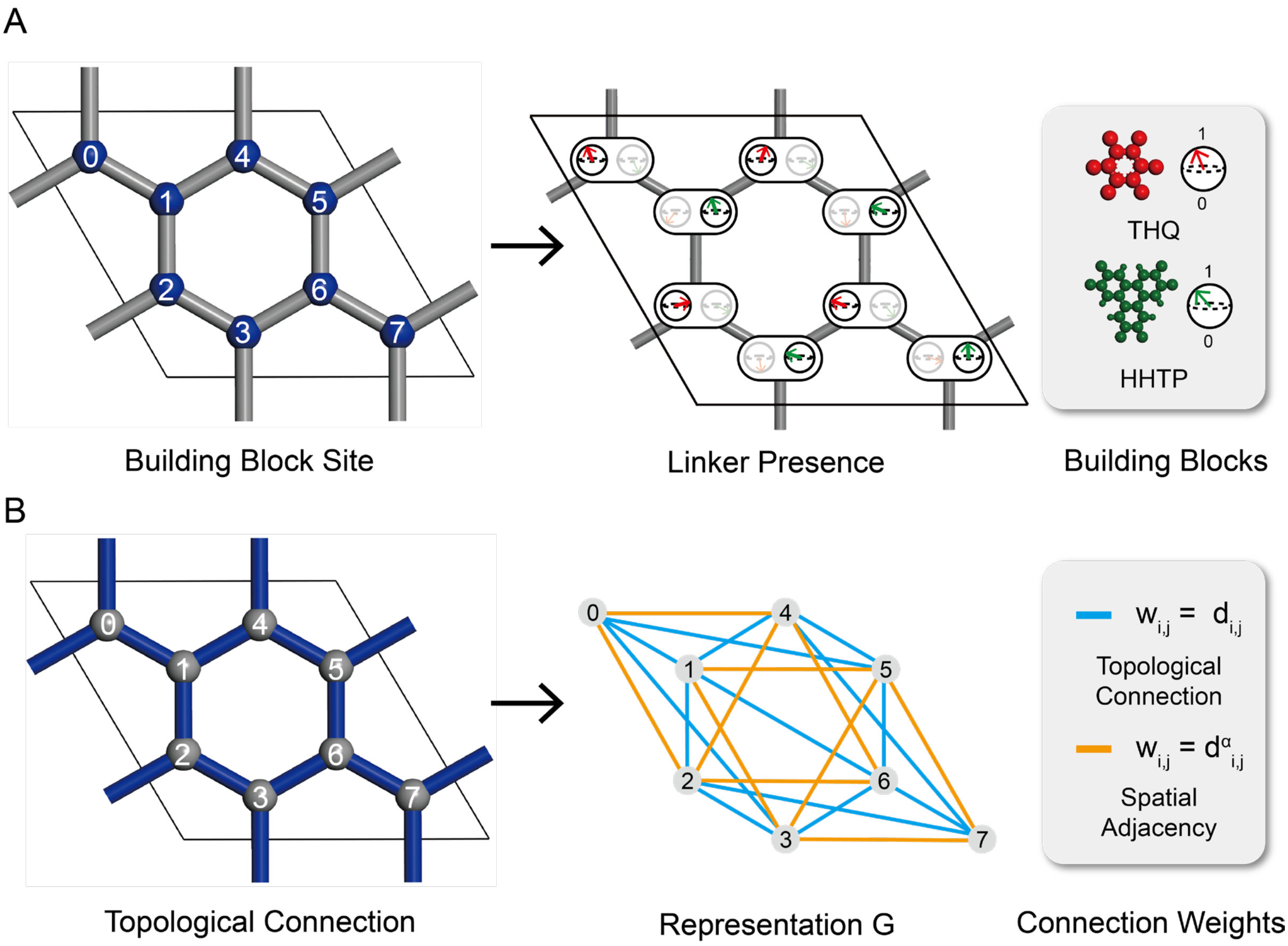

**Fig. 2**. Mapping of the geometrical configuration of Cu-THQ-HHTP MOF with a hcb topology. **(A)** Qubit mapping with two linker candidates in the building block site (blue) with distinct numbering. Two representative qubits (THQ in red and HHTP in green) are allocated in each building block site. Each qubit, representing a single building block type, indicates a linker presence (1) and a linker vacancy (0) in a given site. **(B)** Framework mapping of edges, $(i, j)$, (blue) into a graphical representation, $G(i, j, w_{i,j})$. Each connection is weighted by $w_{i,j}$, which quantifies the strength of either the direct topological connection (light blue) or the spatial adjacency (yellow).

Next, the interactions between these qubits are described by a graph-based framework representation, denoted as $G(i, j, w_{i,j})$, with $G$ symbolizing the connectivity of the MOF framework. Indices $i$ and $j$ represent distinct linker sites within a unit cell, with each ordered pair $(i, j)$ defining an edge. Edges represent either direct topological connections (i.e. linker sites that are connected to one another directly by an edge) or spatial adjacency (i.e. linker sites that are not directly bonded but positioned as the next-nearest neighbors), allowing for indirect interactions. In this paper, the spatial adjacency is limited to the second-closest edges, thereby balancing the computational cost. The graph-based framework representation looks similar to the actual material topology as shown in Figure 2B but it provides additional information about how connected linker sites would influence each other.

The distinction between topological connection and spatial adjacency is achieved by introducing a connection weight, $w_{i,j}$, defined as $w_{i,j} = {d_{i,j}}^{\alpha}$. Here, $d_{i,j}$, denotes the spatial distance (in the unit of Angstroms) between nodes $i$ and $j$, while the sensitivity parameter, $\alpha$, accounts for the type of connection. Specifically, $\alpha$ varies based on whether the connection is a topological connection (first-nearest neighbor, $\alpha = 1$) or a spatial adjacency (second-nearest neighbor, $0 \leq \alpha < 1$), as shown in Equation 1. The connection weight ensures that both topologically connected and spatially adjacent edges contribute to the framework design, with their influence modulated by $d_{i,j}$ and $\alpha$. The reason $\alpha$ varies for second-nearest connections is that linker lengths and spatial distances, $d_{i,j}$, differ depending on the topology type and the set of linker candidates. Fixing $\alpha$ at a single value (e.g. $\alpha = 0.5$) alters the frequency at which the lowest Hamiltonian solution is observed, thereby influencing the final probability distribution of the lowest Hamiltonian, as demonstrated in supplementary table S1. Therefore, a comparative analysis by varying $\alpha$ is necessary to ensure an appropriate selection of this parameter. For spatial adjacency, a lower $\alpha$ reduces the weight, reflecting the diminished impact of non-bonded interactions compared to direct bonds. This formulation enables $w_{i,j}$ to capture varying influence of spatial distance based on the relative importance of connection types. This approach is broadly applicable, as $G(i, j, w_{i,j})$ can be customized to reflect the unique connectivity and spatial relationships of different topologies.

$$G(i, j, w_{i,j}) = \begin{cases} G(i, j, d_{i,j}), & \textit{topological connection } (\alpha = 1) \\ G(i, j, d_{i,j}^{\alpha}), & \textit{spatial adjacency } (0 \leq \alpha < 1) \end{cases} \tag{1}$$

**Reticular Framework Topology-inspired Hamiltonian Design**

With the graph-based framework representation in place, we can next develop a simplified Hamiltonian cost function that can use basic qubit operations to differentiate between the high and the low energy states. We note that this Hamiltonian is different from the actual Hamiltonian of the many-body Schrödinger Equation, which is computationally expensive and cannot be mapped onto the existing quantum computing hardware.

In designing the model Hamiltonian for MTV porous materials, we developed a cost function composed of three key terms: (1) ratio cost, (2) occupancy cost, and (3) balance cost terms as shown in Equation 2 and in Figure 3. Each term addresses a critical aspect of the MTV materials design, ensuring that the Hamiltonian accurately reflects the desired constraints and stability of the material structure within the predefined connectivity framework, $G$.

$$H(q) = H_{ratio}(q) + H_{occupancy}(q) + H_{balance}(q) = \sum_{t \in \{A,B,C,\dots\}} \left( \sum_{i=0}^{i=N_i-1} q_i^t - n_t \right)^2 + \sum_{i=0}^{i=N_i-1} \left( \sum_{t \in \{A,B,C,\dots\}} q_i^t - 1 \right)^2 + \sum_{G \in (i,j,w_{i,j})} w_{i.j} (L(q, G) - \bar{L})^2 \tag{2}$$

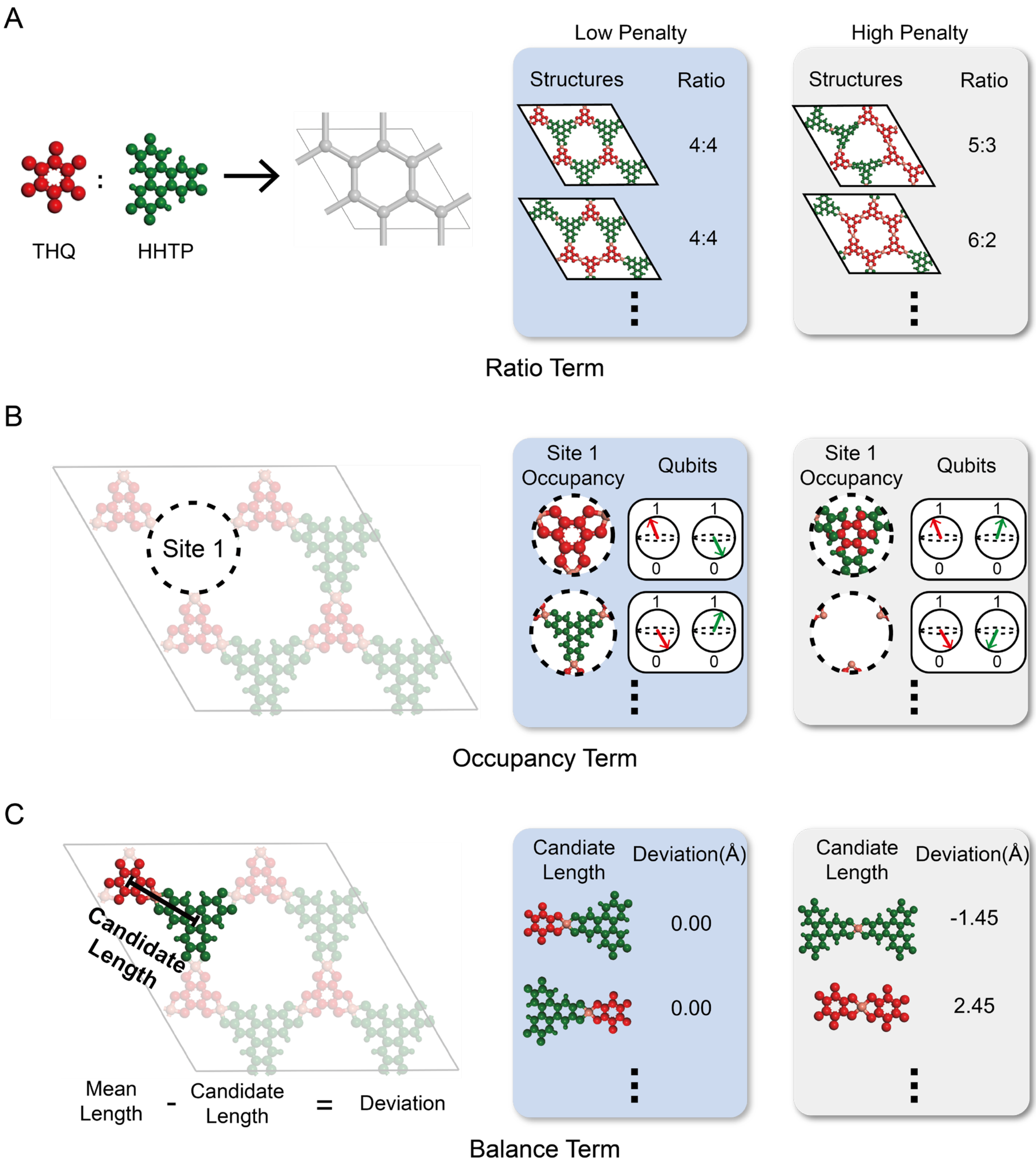

**Fig. 3**. Figure representation of the Hamiltonian cost terms. Blue boxes show examples of low penalties according to our terms and the grey boxes show examples of high penalties according to our terms **(A)** Concept of ratio cost term with 4:4 ratio of each linker (THQ and HHTP). Examples show structures with correct and incorrect ratios **(B)** Concept of occupancy cost term with only one linker, THQ or HHTP, occupying a linker site 1. Examples show qubits representing linker occupancy and linker with correct (Example in blue boxes occupied by one linker) and incorrect (Example in grey boxes occupied by two linkers or vacant site) occupancy **(C)** Concept of balance cost term measuring the deviation of a candidate length $L(q, G)$ with mean edge length, $\bar{L}$. The blue example box shows a well-ordered linker distribution, which has minimal deviation from the mean edge length of 7.29 Å, while the gray example box shows a polarized linker distribution, where HHTP linkers cluster on one side and THQ linkers cluster on the other, causing a high deviation from the mean edge length.

The ratio cost term, $H_{ratio}(q)$, enforces the user-desired ratio of different linker types within the MOF framework, denoted by $n_t$, where $n$ represents the desired proportions and $t$ represents the type of linker. For instance, consider a unit cell of eight linker sites ($\mathrm{N}_i = 8$) in which two linker candidates, THQ and HHTP linkers, are arranged in a 1:1 ratio to form the compound, Cu-THQ-HHTP. Here, Cu-THQ-HHTP indicates all possible configurations composed of Cu metal coordinated with THQ and HHTP likers based on hcb topology, rather than referring exclusively to the experimentally reported structure such as $Cu_3$(HHTP)(THQ)[14]. To simulate all possible configurations while maintaining the desired 1:1 ratio, the number of each linker type, $n_t$, is set to 4 (i.e. $n_{THQ} = 4$, $n_{HHTP} = 4$) as there are eight linker sites in the defined unit cell, which is illustrated in Figure 3A. Consequently, the ratio cost of Cu-THQ-HHTP is represented in Equation 3 below.

$$H_{ratio}(q) = \sum_{t \in \{THQ, HHTP\}} \left( \sum_{i=0}^{7} q^t{}_i - n_t \right)^2$$

$$= (q_0{}^{THQ} + q_1{}^{THQ} + \ldots + q_7{}^{THQ} - 4)^2 + (q_0{}^{HHTP} + q_1{}^{HHTP} + \ldots + q_7{}^{HHTP} - 4)^2 \qquad (3)$$

By penalizing deviations away from the correct linker ratio, this cost function helps optimize configurations that adhere to the material's compositional constraints (which is 1:1 ratio in this example).

Next, the occupancy cost term, $H_{occupancy}(q)$, is introduced to ensure that each linker site is occupied by exactly one linker. Given the fixed topology of MTV porous materials defined in $G$, where each node (i.e. linker site) must be filled by a unique linker to avoid overlapping or empty positions, this term penalizes configurations with either multiple linkers at the same single site or no linker at all. For example, as shown in Equation 4 and Figure 3B, the occupancy cost of Cu-THQ-HHTP penalizes any instance where a linker site does not meet this condition, preventing overlapping linkers or vacant sites. As a result, this constraint prevents non-physical chemical configurations from entering the solution space.

$$H_{occupancy}(q) = \sum_{i=0}^{7}(\sum_{t\in\{THQ,HHTP\}} {q^t}_i - 1)^2 = ({q_0}^{THQ} + {q_0}^{HHTP} - 1)^2 + \cdots + ({q_7}^{THQ} + {q_7}^{HHTP} - 1)^2 \quad (4)$$

Finally, the balance cost term, $H_{balance}(q)$, is constructed to maintain a spatially balanced arrangement of building blocks within the topology of MTV porous materials. Previous experimental studies on MTV MOFs and COFs have shown that well-ordered linker distributions contribute to structural stability(13-18, 27), as they minimize geometric strain and prevent excessive aggregation of specific linker types, which could lead to local distortions (fig. S1). Therefore, the balance cost term is designed to promote a uniform spatial distribution of linkers with varying lengths by minimizing deviations of individual edge lengths, $L(q, G)$, from a mean edge length, $\bar{L}$. In this term, $L(q, G)$ represents the individual length (in Angstroms) of each edge $(i, j)$ and it is the sum of characteristic lengths occupying linker sites $i$ and $j$, as defined in Equation 5

$$\mathrm{L(q,G)} = \sum_{t_1\in\{A,B,\ldots\}}\sum_{t_2\in\{A,B,\ldots\}}(l^{t_1}q_i^{t_1} + l^{t_2}q_j^{t_2}) \quad (5)$$

where $l^{t_1}$ and $l^{t_2}$ are the characteristic lengths of linker types that belong to $t_1$ and $t_2$, and $q_i^{t_1}$ and $q_j^{t_2}$ indicate the presence of linkers at sites $i$ and $j$, respectively. The characteristic length represents the length of each linker within the framework. For instance, tritopic linkers such as THQ, which form three connections with metal clusters, have a characteristic length equivalent to the radius of the circle that links these points resulting in 2.42 Å and 4.87 Å for THQ and HHTP, respectively (table S2). In contrast, ditopic linkers such as BDC (benzene dicarboxylate), which connect metal clusters linearly, have a characteristic length of 2.87 Å that spans half the entire distance between the connection points (table S2). By incorporating these geometric considerations, in Cu-THQ-HHTP, for instance, the edge length for $G(0,1,3)$ is calculated as:

$$L\big(q, G(0,1,3)\big) = \sum_{t_1\in\{THQ,HHTP\}}\sum_{t_2\in\{THQ,HHTP\}}\big(l^{t_1}q_i^{t_1} + l^{t_2}q_j^{t_2}\big)$$

$$= 2 \times (2.42{q_0}^{THQ} + 4.87{q_0}^{HHTP} + 2.42{q_1}^{THQ} + 4.87{q_1}^{HHTP}) \quad (6)$$

The mean edge length, $\bar{L}$, serves as a stable reference to minimize deviations across edge lengths. Since the ratio of different linker types is predefined by $n_t$ in the ratio cost term, $\bar{L}$ remains constant across all linker arrangements. This provides a consistent target for minimizing deviations in $L(q,G)$, promoting a uniform linker arrangement to prevent structural distortions (Fig. 3C). The mean edge length, $\bar{L}$, is defined in Equation 7:

$$\bar{L} = \frac{1}{|G|}\sum_{G\in(i,j,w_{i,j})} L(q,G) \quad (7)$$

where $|G|$ is the total number of edges in $G$. For example, in a unit cell of Cu-THQ-HHTP, $|G|$ is 24 and $\bar{L}$ is 7.29 Å. The balance cost term is then expressed as:

$$H_{balance}(q) = \sum_{G \in (i,j,w_{i,j})} w_{i,j}(L(q, \mathrm{G}) - 7.29)^2 \qquad (8)$$

This term is weighted by $w_{i,j}$, which quantifies the strength of the connection between nodes $i$ and $j$, either through direct topological bonds or spatial adjacency. By incorporating these weights based on the connection type (e.g. assigning stronger weights to topologically bonded pairs and weaker weights to spatially adjacent pairs), the balance cost accurately reflects the geometrically optimal linker arrangements, which is illustrated in Figure 3C. This term also can be understood by promoting a uniform pore distribution within the final structure (fig. S1).

**Reproducibility of the Hamiltonian Model to Real MTV Reticular Frameworks**

The solution to the MTV material design problem corresponds to finding the ground state of the Hamiltonian, $H(q)$, which minimizes the ratio, occupancy, and balance costs across the predefined graph-based framework, $G$. To test our Hamiltonian model and validate its ability to reproduce experimentally reported MTV porous materials, a variational quantum circuit was constructed and executed using the Sampling VQE algorithm in IBM Qiskit (26).

The VQE is a hybrid quantum-classical algorithm that approximates the ground state of a given Hamiltonian by iteratively optimizing a parameterized quantum circuit. In variational quantum algorithms, an ansatz refers to a structure for a parameterized quantum circuit designed to generate trial quantum states (28). The ansatz defines a sequence of unitary operations to manipulate quantum states of qubits initialized in a computational reference state. These unitary operations consist of ansatz parameters, θ, and these are the one being iteratively optimized via VQE algorithm to approximate the ground state of the Hamiltonian (Fig. 4A(i)). The ansatz parameters for the reference states are randomly initialized from the range -2π to 2π. Upon convergence of the VQE process, the trial quantum state approximates the system's ground-state wavefunction (28).

For this study, we used a Two Local ansatz which consists of parameterized single-qubit Ry rotations, controlled-Z (CZ) gates in a linear entanglement structure, and additional single-qubit Ry rotations (Fig. 4A). The choice of the ansatz is based on its simple yet effective framework for exploring the solution space with minimal circuit complexity (29, 30), which is critical for this study, as the primary objective is to confirm the viability of the Hamiltonian model rather than optimize for larger, more complex systems. The circuit depth was kept minimal by setting the number of repetitions to 1 (one layer of entangling gates), resulting in a total number of circuit parameters to twice the number of qubits, $2|t| \cdot N_i$. For more details on the parameter setup for the ansatz, readers might refer to the method section.

Once the quantum circuit is prepared, the variational quantum algorithm optimizes the ansatz parameters, θ, to minimize the expectation value of the Hamiltonian. Our Hamiltonian model, $H(q)$, is diagonal in the computational basis as it involves only classical binary variables representing linker presence and their associated costs. In a diagonal Hamiltonian, the eigenvalues correspond directly to the measurement outcomes of the quantum circuit, greatly simplifying the evaluation of the expectation value, $E(\theta)$(31). The Sampling VQE algorithm, a variant of the VQE, is particularly suited for such diagonal Hamiltonians. Unlike the standard VQE, which computes the expectation value of the Hamiltonian using exact state vectors or an simulator, Sampling VQE evaluates $E(\theta)$ by sampling measurement outcomes of the trial states prepared by the quantum circuit (26). Sampling refers to the process of repeatedly running the quantum circuit to measure the outcomes. Each run of the circuit constitutes a shot, and the resulting probability distribution is derived from the frequencies of these measurement outcomes across the total number of shots (Fig. 4B). The sampling mimics the behavior of near-term quantum hardware, where noise and finite sampling inherently limit the precision of the measured outcomes. This sampling process is iteratively performed to calculate $E(\theta)$. Specifically, if $x$ represents the binary measurement outcome of the qubits, the expectation value is evaluated as:

$$E(\theta) = \sum_x P(x|\theta)H(x) \tag{9}$$

where $P(x|\theta)$ is the probability of measuring the state $x$, and $H(x)$ is the value of the Hamiltonian for that state (Fig. 4B). The variational parameters $\theta$ collectively represent a set of tunable parameters applied to all qubits in the ansatz circuit, creating a probabilistic distribution over multiple quantum states. For example, consider a specific set of parameters, $\theta_A$, prepared for an $n$-qubit system. It is not the case that $\theta_A$ deterministically encodes only one of the $2^n$ possible states. Rather, $\theta_A$ determines the probability amplitudes of all $2^n$ states, and each measurement collapses the quantum state into one of these possible configurations based on the probability distribution induced by $\theta_A$. During optimization, the classical optimizer then minimizes $E(\theta)$ by updating the full set of $\theta$ based on stochastic gradient approximation. The updated parameters are then applied to all qubits in the next iteration to generate a new trial quantum state. This iterative optimization process continues until the optimization converges or the desired number of iterations for parameter updates is reached. Given its computational efficiency and similarity with realistic quantum measurements, the sampling VQE algorithm was used to validate our Hamiltonian model by determining whether its ground state corresponds to experimentally reported MTV porous material structures.

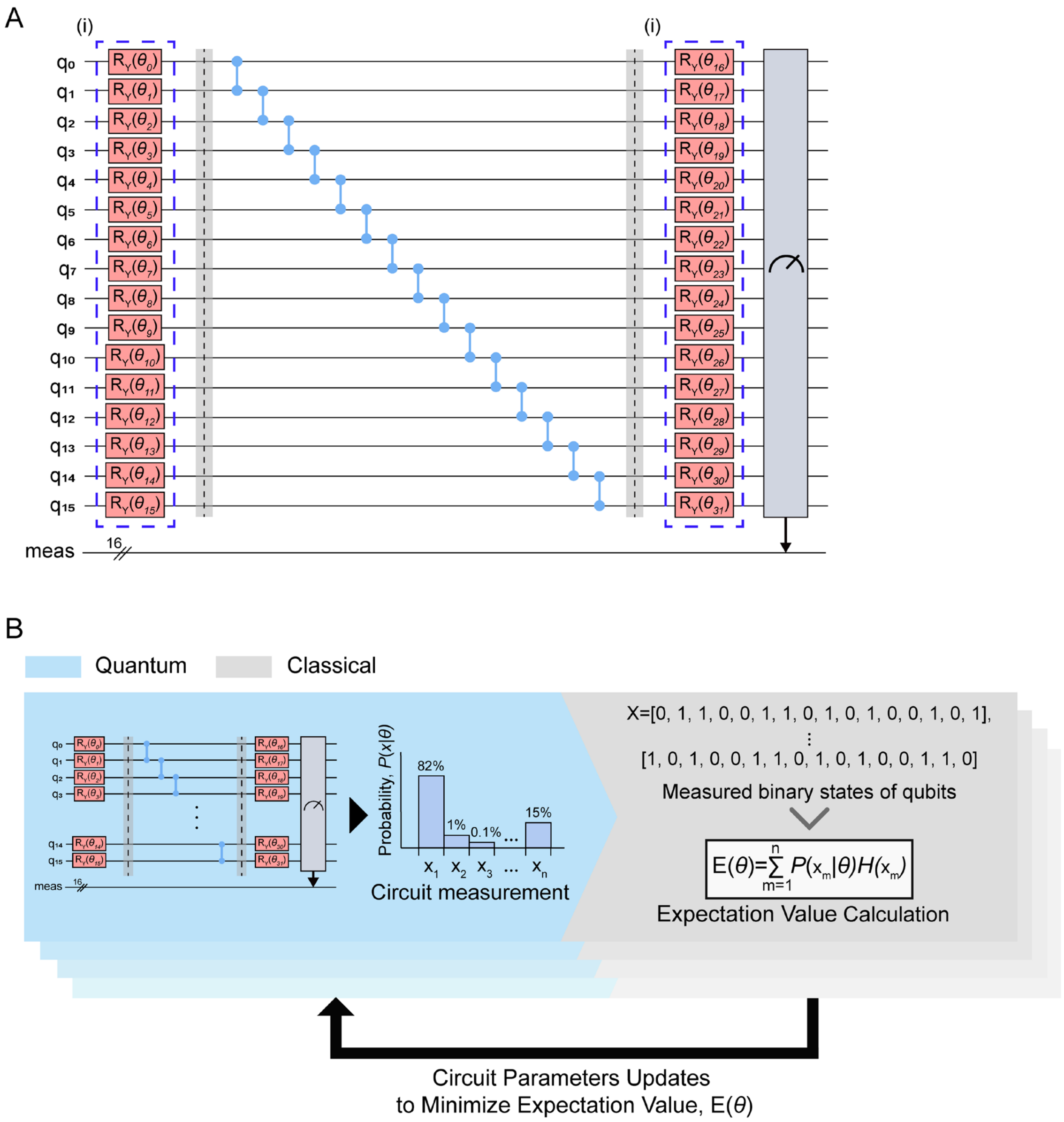

**Fig. 4. (A)** Quantum circuit of Cu-THQ-HHTP within a unit cell consisting of an eight-linker-site system, based on Two Local ansatz. **(i)** Quantum circuit parameters, θ, are highlighted with blue dotted line boxes. **(B)** An overall process of sampling VQE algorithm. The sampling process generates over $n$ possible measured binary state, $x$ (i.e. $x_1, x_2, \ldots, x_n$). The set of measured binary states, $X$, along with their corresponding probabilities, $P(X|\theta)$, and Hamiltonian values, $H(X)$, are used to calculated the expectation value, $E(\theta)$.

Based on the choice of ansatz and quantum algorithm, our Hamiltonian model was applied to simulate four experimentally known MTV porous materials: Cu-THQ-HHTP, Py-MV-DPA-COF, MUF-7, and SIOC-COF2. These structures were selected for their structural diversity, with Cu-THQ-HHTP and Py-MV-DPA-COF representing 2D MOF and 2D COF structures based on the hcb topology (14, 32), MUF-7 as a 3D MOF based on ith-d topology (13), and SIOC-COF2 as a 2D COF based on the kgm topology (15) (fig. S2). This structural diversity provides an opportunity to assess the adaptability of our model across a range of reticular frameworks. All of these topologies were translated into graph-based representations, $G(i, j, w_{i,j})$, under specific assumptions regarding the unit cell size, $N_i$, to ensure that the number of qubits does not exceed 20, which is the upper limit for the computational resources available. Readers may refer to supplementary materials note S1 and table S3 for detailed computational methods related to circuit construction and simulation.

Figure 5 shows the final probability distributions of each structure, obtained from the Sampling VQE simulations by averaging 128 independent runs using a fixed sampler seed. This approach ensures a more stable estimate by mitigating fluctuations across individual runs. Specifically, it contains only the probability values associated with the top six lowest Hamiltonians, while complete probability distributions are provided in supplementary figure S3. The ground states (i.e. lowest Hamiltonian values) of the developed Hamiltonian have all correctly reproduced the experimental configurations with the highest probabilities, demonstrating that (a) our constructed Hamiltonian is a reasonable one and (b) the quantum computing algorithm correctly identifies the optimal values. SIOC-COF2 resulted in the highest ground state probability at 30.9%, MUF-7 was the second highest as 25.5%, Cu-THQ-HHTP and Py-MV-DBA-COF resulted in 16.3% and 13.5%, respectively. Differences of the ground state probability among structures can be understood by the complexity of the system such as number of qubits and connection weights which are associated with the choice of target topology and linker types. SIOC-COF2 and MUF-7 involve six linker sites in the graph $G$, with two linker candidates for each site, translating to 12 qubits and 24 circuit parameters in their quantum circuits. In contrast, Cu-THQ-HHTP and Py-MV-DBA-COF involve eight linker sites with two linker candidates, requiring 16 qubits and 32 circuit parameters. The increased circuit complexity and larger Hilbert space result in a more dispersed probability distribution (fig. S3), thereby lowering the probability of the ground state configuration.

In addition, the highest ground state probability of SIOC-COF2 can also be attributed to the simplification of its connection weight, $w_{i,j}$ due to absence of secondary connections. In the defined unit cell of SIOC-COF2, all linker sites are topologically connected, resulting in α=1 for all edges in $G_{kgm}$ (fig. S2C). This uniformity makes the spatial distance, $d_{i,j}$, the only factor influencing the connection weight, thereby further simplifying the Hamiltonian model. In contrast, the other structures require careful selection of α through the comparative analysis, varying its values from 0 to 1. This analysis was performed using four different settings where α was set to 0.01, 0.1, 0.25, and 0.5 (table S1). However, this limited testing may not sufficient to identify the optimal α value for each structure, especially given their distinct characteristic lengths, $l$, spatial distances, $d_{i,j}$, and unique topologies, $G$. Despite the simplification of quantum circuit design and simulation, the results proved the effectiveness of our Hamiltonian model in reproducing the experimental configurations. As a result, the proposed Hamiltonian model showed the potential extensibility to complex systems such as larger unit cells with many linker candidates and different connectivity.

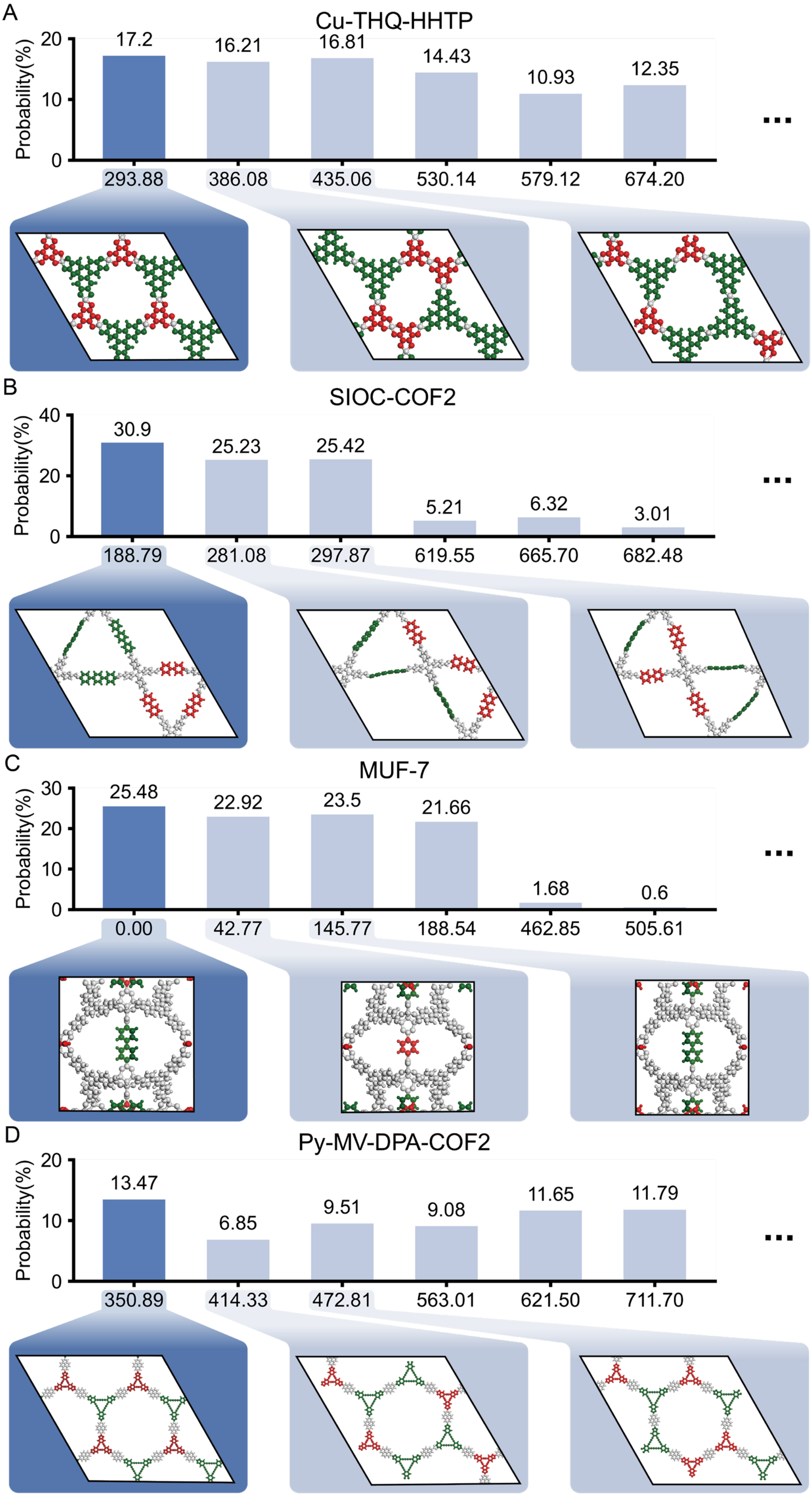

**Fig. 5.** Final probability distribution of four experimentally known MTV structures. For clarity, only the probabilities corresponding to the six lowest Hamiltonian are shown. The structure of three low Hamiltonian are shown and the structure with the lowest Hamiltonian and highest probability is marked with dark blue. Linkers within the structure are marked by their characteristic length, marking shorter linker as red and longer linker as green **(A)** Probability distribution of Cu-THQ-HHTP with its Hamiltonian values and the respective configuration. The lowest Hamiltonian structure (293.88), which corresponds to the experimental structure of Cu-THQ-HHTP, showed the highest probability. **(B)** Probability distribution of SIOC-COF2, where the lowest Hamiltonian (188.79) with the highest probability matches the experimental structure of SIOC-COF2. **(C)** Probability distribution of MUF-7, where the lowest Hamiltonian (0.00) with the highest probability matches the experimental structure of MUF-7. **(D)** Probability distribution of Py-MV-DPA-COF2, where the lowest Hamiltonian value (350.89) with the highest probability matches the experimental structure of Py-MV-DPA-COF2.

Finally, among the four candidate materials, we selected one (SIOC-COF2) with the least number of qubit requirements and performed VQE calculations on IBM's real quantum hardware (ibm_kyiv) to assess its performance and consistency in estimating the ground state energy. Unlike the classical Sampling VQE results, which were based on 300 optimization iterations, the quantum hardware simulation was limited to 50 iterations due to constraints on quantum computational resources. As shown in Figure 6, the expectation values from quantum hardware demonstrated a clear convergence trend with a final expectation value of -1284.6. This trend closely aligns with the results from Qiskit's Estimator primitive using the Aer backend, which models an ideal, noiseless quantum system. The classical simulation results were derived from averaging 128 independent VQE runs using a fixed simulator seed, providing a more stable estimate by accounting for fluctuations in individual runs. The slightly better performance of the classical simulation is expected, as the Aer simulator does not incorporate real hardware errors, whereas quantum hardware is subject to noise and gate errors. Despite these challenges, the quantum results were in good agreement with the classical simulations. Given that the optimal expectation value, $E_{optimal}(\theta) = \sum_x P(x|\theta)H(x)$, where $P(x|\theta) = 1$, is estimated at -4385.9, it is expected that a quantum simulation with a sufficiently large number of iterations (e.g., over 500 iterations) would successfully approach this optimal value. This result further validates the reliability of our Hamiltonian model in estimating the ground-state configuration.

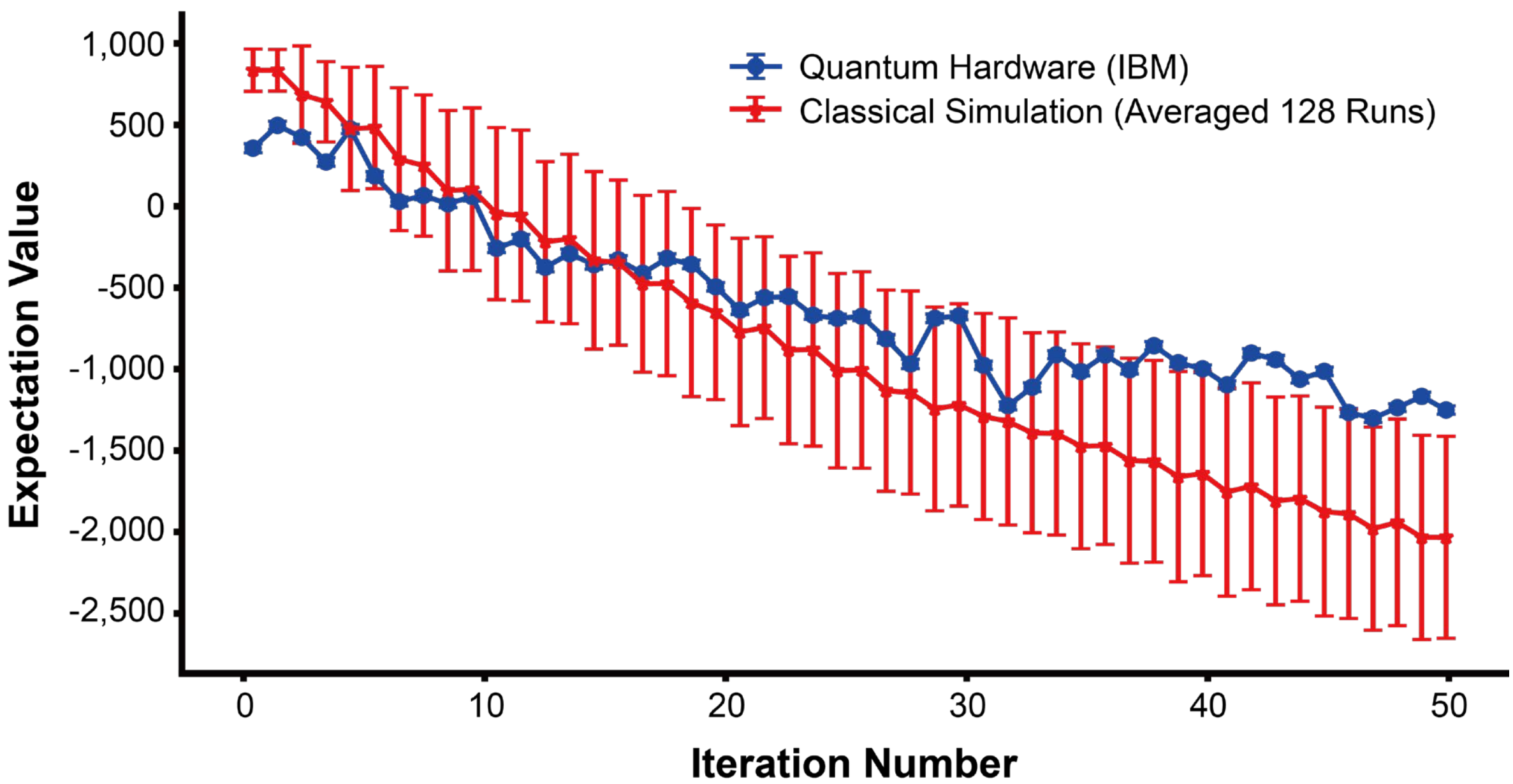

**Fig. 6.** VQE calculation results comparing IBM quantum hardware (ibm_kyiv) with classical simulation. The blue markers represent expectation values obtained from the quantum

hardware, with corresponding error bars indicating standard deviations. The red markers show results from classical simulations, performed using the Estimator primitive from Qiskit Aer. Expectation values were averaged over 128 independent runs using a fixed simulator seed (123), with error bars representing standard deviations. The graph illustrates the optimization process, where expectation values decrease as the number of iteration increases.

**Extensibility of the Hamiltonian Model to Complex MTV Reticular Frameworks**

While our current simulation results demonstrate the reproducibility of experimentally known material configurations for relatively simple porous material unit-cells (i.e. systems requiring fewer than twenty qubits), we want to emphasize the potential extensibility of our quantum computing-based approach for designing complex MTV reticular structures. The complex MTV reticular structures refer to MTV porous materials with intricate spatial arrangements of building blocks that go beyond simple periodicity. These structures arise when the design constraints, such as varying linker ratios and diverse geometric lengths, necessitate the extension of the primitive unit cell to accommodate the required number of linker sites. For example, designing an MTV material based on hcb topology with an arbitrary chosen linker ratio for four distinct linkers requires expanding the primitive two-site unit cell to a seventy-two-linker-site unit cell (Fig. 7). The structure is complex as their nonuniform proportions and varying spatial lengths of the linkers make it challenging to intuitively determine the optimal spatial configuration. This structural complexity arises from the interplay of conflicting chemical and structural factors as the building blocks adapt to the framework formation (33). Chemists are motivated to synthesize these materials due to their synergistic functionalities, which exceed the sum of their individual components (34, 35). However, despite relying on known chemical intuitions in the design of complex reticular structures, accurately predicting whether such materials are even feasible with classical computing becomes increasingly challenging as the number of constituent building blocks grows, leading to an exponential increase in possible configurations.

In our proposed Hamiltonian model, the problem complexity of the MTV material design is influenced by three variables: (1) MTV linker types, $t$, (2) the predefined proportions of MTV linkers, $n_t$, and (3) the number of linker sites in a defined unit cell, $N_i$. Each linker site within the framework can adopt one of the linker types as $t \in \{A, B, C, ...\}$, and the total number of linker sites, $N_i$, governs the size of the framework. Without considering ratio constraints, the problem spans a vast configuration space of $2^{|t| \cdot N_i}$, as each linker site independently takes on one of the binary configurations for the $|t|$ linker types. Once the predefined ratio, $n_t$, is introduced, it significantly reduces the dimensional space by ensuring that only configurations satisfying $\sum_{t \in \{A,B,C,...\}} n_t = N_i$ are valid. Therefore, the reduction in the configuration space due to $n_t$ can be described by the multinomial coefficient:

$$N_{MTV\ config.} = \binom{N_i}{n_A, n_B, n_C, ...} = \frac{N_i!}{n_A!\, n_B! n_C!\ ...} \quad (10)$$

where $N_{MTV\ config.}$ represents the total number of MTV configurations and $n_A, n_B, n_C$... represent the respective counts of each linker type as defined by $n_t$. For instance, Cu-THQ-HHTP with the eight-linker site unit cell consists of $|t| = 2$ linker types (THQ and HHTP), $N_i = 8$, and a user-desired ratio of $\{n_{THQ},\ n_{HHTP}\} = \{4, 4\}$. Its dimensional space reduces from $2^{16} = 65{,}536$ to 70 valid configurations that satisfy the ratio constraint. Although the introduction of the ratio constraints reduces the configuration space, Equation 10 still highlights the exponential increase in design complexity as the number of tunable variables $(t, n_t, N_i)$ increases.

The quantum computing approach based on the proposed Hamiltonian model can provide a significant advantage over classical brute-force methods in addressing this exponential complexity. Figure 7 illustrates the exponential increase in the number of possible MTV structures as the unit cell size expands from the primitive two-linker-site system to the seventy-two-linker-site system for the hcb framework. The primitive unit cell, with two linker types ($|t| = 2$), serves as a simpler case, while larger unit cells such as the seventy-two-linker-site system, incorporate four linker types ($|t| = 4$), significantly increasing the structural complexity. Although the number of qubits required to represent the system scales linearly with the equation $N_{qubits} = 4N_i$ from the eight-linker-site unit cell onward, the number of the MTV structures grows exponentially with the multinomial coefficient (Equation 10). In classical computing, the Hamiltonian must be evaluated individually for every possible configuration, which becomes infeasible for complex systems. For instance, simulating an hcb topology in the seventy-two-linker site unit cell with four types of linkers would require a classical computer to simulate approximately $7.45 \times 10^{34}$ configurations which are astronomical (Fig. 7). The computational resources and time needed to evaluate each structure one by one would ultimately render the problem intractable. On the other hand, the proposed VQE-based quantum algorithm can efficiently explore this vast search space and identify optimal configurations with a single measurement based on the principles of quantum mechanics. While the limitations of quantum resources in Noisy Intermediate-Scale Quantum (NISQ) technology currently restrict our ability to simulate highly complex MTV materials, we believe the proposed Hamiltonian model could enable the discovery and design of such materials that are beyond the reach of classical methods.

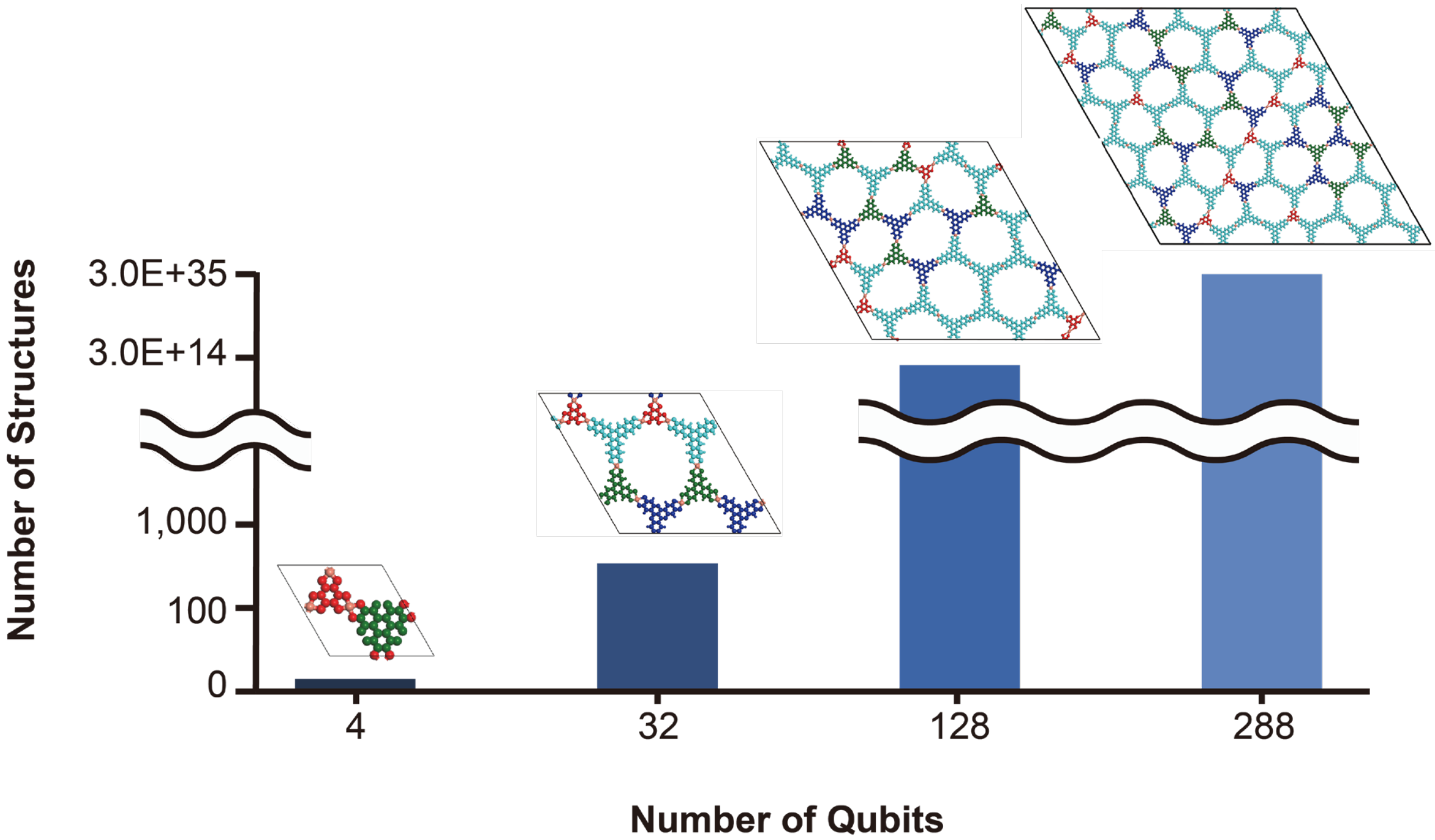

**Fig. 7.** The exponential growth of possible MTV porous material configurations as the number of qubits increases. The x-axis represents the number of qubits required to encode MTV porous materials for unit cells containing two, eight, thirty-two, and seventy-two linker sites, corresponding to 4, 32, 128, and 288 qubits, respectively. The y-axis shows the total number of possible MTV porous structures that can be computed based on the corresponding qubits. The molecular structures of THQ, HHTP, HHTT (2,3,7,8,12,13-hexahydroxytetraazanaphthotetraphene), and HHTN (2,3,8,9,14,15-decahydronaphthalene) linkers are highlighted in red, green, blue, and cyan colors, respectively. Representative structures for each unit cell configuration are shown as inset images above the corresponding bars in the graph.

## Discussion

In this work, we developed the Hamiltonian model designed for gate-based quantum computing to design MTV porous materials with desired building block combinations. Inspired by geometrical intuitions from experimental MTV structures, the proposed model is designed as a coarse-grained model by embedding compositional, structural and balance constraints directly into the Hamiltonian. This method enables efficient optimization of MTV configurations, allowing the identification of optimal arrangements of building blocks that satisfy predefined design criteria. Our model introduced a 2D graph-based topology representation, $G(i, j, w_{i,j})$, incorporating connection weight, $w_{i,j}$ to account for spatial distance, $d_{i,j}$ , and connection type, $\alpha$. This approach captures the relative contributions of individual building blocks to the overall structure and can be customized for various topologies, making it broadly applicable. To validate the model, we implemented it on a variational quantum circuit using the sampling VQE algorithm in IBM Qiskit. Simulations on four experimentally known MTV porous materials, Cu-THQ-HHTP, Py-MV-DBA-COF, MUF-7, and SIOC-COF2, successfully identified the ground-state Hamiltonian configurations, aligning with experimental results and demonstrating the potential of our approach in accurately simulating complex MTV structures. To assess our Hamiltonian model on real quantum hardware, we performed VQE calculations on a 12-qubit system for SIOC-COF2 using ibm_kyiv processor. As the structure with the lowest qubit requirement, SIOC-COF2 served as a benchmark. The expectation values obtained from the quantum hardware exhibited a clear convergence trend, closely aligning with the results from classical simulations using IBM Qiskit's Aer simulator. The slightly better performance of the classical simulation is expected, as the Aer simulator assumes an ideal, noiseless quantum system, while real quantum hardware is inherently affected by noise and gate errors. Nevertheless, despite these challenges, the quantum results remained highly consistent with the classical simulations, demonstrating the reliability of our model and highlighting the potential of quantum algorithms for MTV material design.

However, we acknowledge that the proposed Hamiltonian model is primarily based on topological and geometrical approximations, capturing only a fraction of the complexities inherent in reticular frameworks. Molecular science, which is governed by the dynamics of electrons and atomic nuclei and their interactions with electromagnetic fields, often requires detailed quantum mechanical models for accurate predictions[36], but near-term quantum devices currently impose computational limitations. Moreover, we do not aim to make claim on the difficulty of actually synthesizing these complex MTV porous materials, which might have other difficulties (e.g. diffusion limitations, sub-optimal experimental synthesis conditions). Despite this, the coarse-grained approach provides an essential first step towards exploring the vast design space of MTV materials, utilizing quantum computing's ability to represent exponentially large wave functions with a linear scaling of qubits. As demonstrated, the dimensional space of MTV configurations grows exponentially with the number of linker types, proportions, and sites (Equation 10). Classical brute-force methods ineffectively navigate such a vast combinatorial landscape due to the need for individual evaluations of every configuration. In contrast, our quantum algorithm can efficiently explore their high-dimensional design space, identifying ground-state configurations through a single quantum measurement and circumventing the exhaustive calculations required by classical methods. Looking ahead, we believe the advancements in quantum hardware and algorithms could further extend the applicability of our Hamiltonian model for the design of increasingly complex MTV materials. This work establishes a foundation of quantum computing to design next-generation MTV porous materials with unparalleled efficiency.

## Materials and Methods

### Sampling VQE Calculations with Classical Simulator

All classical simulations in this study were conducted using IBM Qiskit modules (26). To optimize the circuit parameters, the SPSA (Simultaneous Perturbation Stochastic Approximation) optimizer was employed, as it is well-suited for noisy and resource-constrained quantum simulations. The parameters were updated for 300 iterations based on the SPSA optimizer. The quantum circuit was executed using a Qiskit Sampler primitive and each simulation was performed using 1024 measurement shots, ensuring statistically significant sampling for the corresponding probability distributions. The Sampling VQE algorithm was executed for 128 independent iterations against the final optimized set of circuit parameters with a fixed sampler seed of 123, each producing a unique probability distribution over the possible structures. To compute the final probability distribution, the probabilities associated with each structure across all 128 runs were averaged and normalized. This averaging process accounts for fluctuations in individual runs and provides a more accurate representation of the likelihood of each structure. By aggregating the results in this manner, the final probability distribution reflects the most probable structural configurations predicted by the Sampling VQE simulations under the given Hamiltonian model.

To determine the optimal $\alpha$ values for each simulated structure, the comparative analysis was conducted using four different settings, with $\alpha$ set to 0.01, 0.1, 0.25, and 0.5 (table S1). The $\alpha$ value was chosen based on the condition that maximizes the occurrences of the lowest Hamiltonian solution with the highest probability (table S1).

## VQE calculations with Quantum Hardware and Classical Simulator

Our Hamiltonian model was used to transform the weighted graph representation of the MTV porous material into a quadratic unconstrained binary optimization (QUBO) problem. The QUBO formulation encodes the linker configurations and their associated energy costs as binary decision variables. These binary variables, denoted as $q$, represent the presence ($q = 1$) or absence ($q = 0$) of a specific linker at a given site. The QUBO problem was then mapped onto an Ising Hamiltonian, $H(q)$, where $H(q)$ defines the total energy of the system based on the selected linker configurations $q$. This Hamiltonian serves as the objective function for quantum simulation, where the goal is to find the ground state $q$ that minimizes $H(q)$.

For the VQE calculation from quantum hardware, the quantum circuit was constructed using a TwoLocal ansatz using a TwoLocal ansatz, with Ry rotation gates for single-qubit operations and CZ entangling gates arranged in a linear topology. The circuit was optimized using Qiskit's transpiler with optimization level 3, ensuring that the compiled circuit adhered to the native gate set of the ibm_kyiv quantum processor (fig. S4). The transpiled circuit was then used for quantum measurement and the corresponding qubit operator was adjusted to align with the optimized circuit layout. The quantum simulation was executed on IBM's ibm_kyiv backend through Qiskit Runtime, which provides cloud-based access to real quantum hardware. The execution involved parameter optimization using the VQE algorithm. The VQE process iteratively updated the variational parameters, $\theta$, by minimizing the expectation value of $H(q)$. Each optimization run was submitted as individual job and corresponding expectation value was obtained using Qiskit's EstimatorV2 primitive.

For the VQE calculations using a classical simulator, we employed IBM Qiskit's Aer simulator, which provides an ideal, noiseless simulation of quantum circuits. The same TwoLocal ansatz used for quantum hardware was used for the classical simulations, ensuring a direct comparison between the two approaches. The classical simulation was executed using the Estimator primitive in Qiskit Aer, with 128 independent VQE simulations, each using a fixed Aer simulator seed (123) to maintain controlled simulator behavior.

Both quantum and classical simulations used the same initial parameters, generated with a fixed seed (10568) to ensure consistency in the optimization process. Optimization was performed using the SPSA optimizer with a learning rate of 0.00009, a perturbation scale of 0.09, and one resampling per iteration to balance accuracy and computational cost. While iteration 1 used the fixed initial parameters, subsequent iterations (2 to 50) were initialized from the optimized parameters of the preceding iteration, ensuring a continuous optimization process.

**Construction of MTV Porous Materials**

We explored the hypothetical configuration of MTV porous materials using the porous materials generation kit, PORMAKE (18). PORMAKE utilizes a top-down approach in constructing a porous material when given target topologies and building blocks. An additional building block data set was added to the program to model the experimental MTV porous materials.

## Acknowledgments

We acknowledge continuous support from the IBM Quantum team. **Funding:** We thank the National Research Foundation of Korea (Project Number RS-2024-00337004, RS-2024-00451160) for the financial support. **Author contributions:** J.K. conceived the idea and supervised the overall project. S.K. designed the Hamiltonian model, performed sampling VQE and VQE simulations, and conducted data analysis. Y.K. generated porous material structures and performed data analysis. All authors discussed the results and provided comments on the manuscript. **Competing interests:** The authors declare no competing interests. **Data and materials availability:** All data needed to evaluate the conclusions in the paper are present in the paper and/or the Supplementary Materials. Additional data and code related to this paper may be found in https://github.com/shinyoung3/QC-MTV and requested from the authors.

# Supplementary Materials for

## Quantum Computing Based Design of Multivariate Porous Materials

Shinyoung Kang[1], Younghun Kim[1], Jihan Kim[1]*

*Corresponding author. Email: jihankim@kaist.ac.kr (J. K.)

**This PDF file includes:**

Figs. S1 to S4

Note S1

Table S1 to S3

References 37, 38

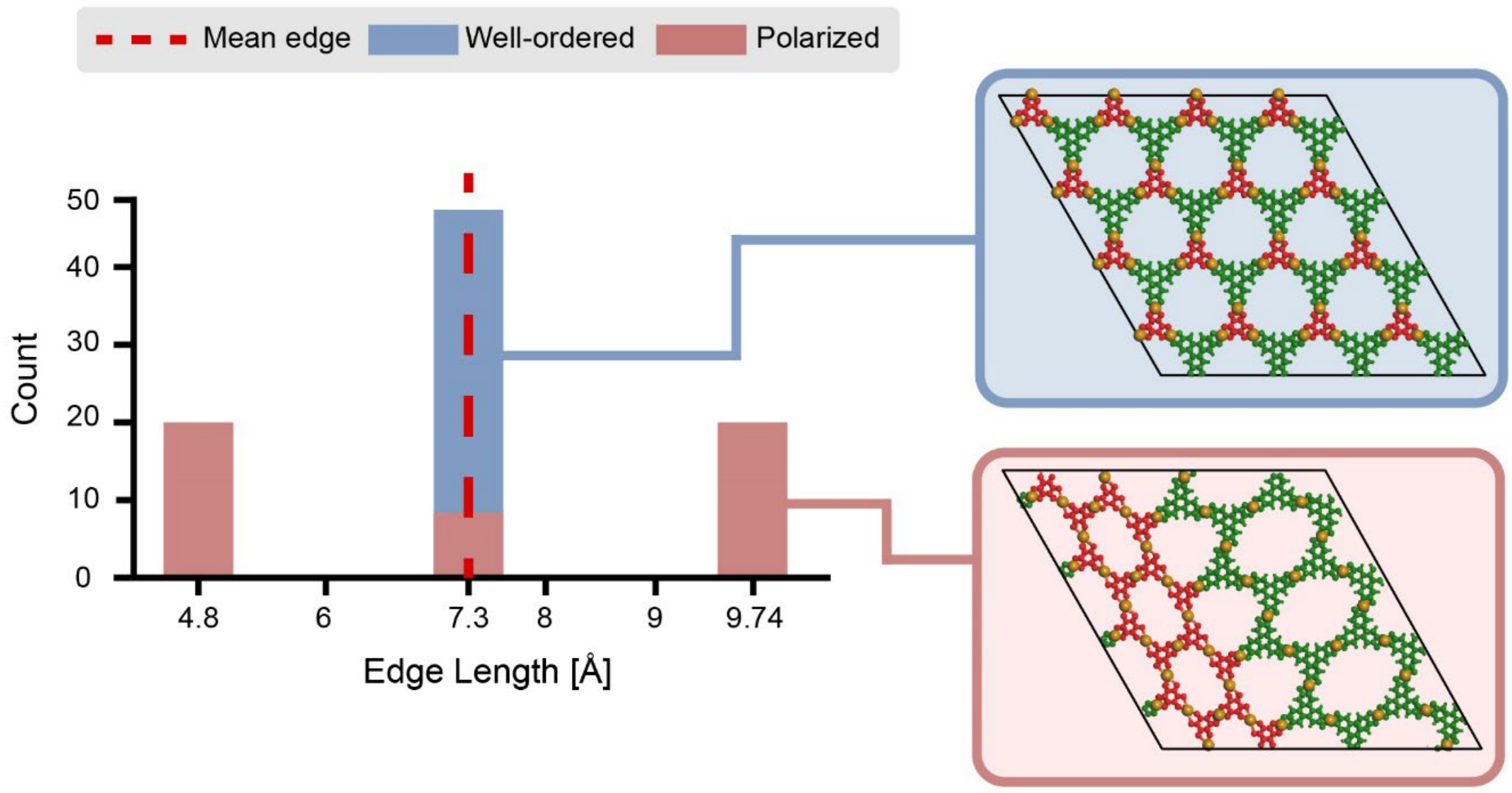

**Fig. S1. Pore distribution of Cu-THQ-HHTP structures.** Distribution of edge lengths for two different Cu-THQ-HHTP configurations. The mean edge length is indicated by a red dotted line, while the edge length distributions for the well-ordered and polarized linker arrangements are shown in blue and pink, respectively. The well-ordered linker arrangement exhibits a narrow distribution of edge lengths, resulting in a uniform pore structure. In contrast, the polarized linker arrangement, where THQ linkers cluster on one side and HHTP linkers cluster on the other, yields a broader edge length range (4.8 to 9.74 Å). This variation leads to local distortions in the pore structure due to bond lengthening between the metal nodes and linkers.

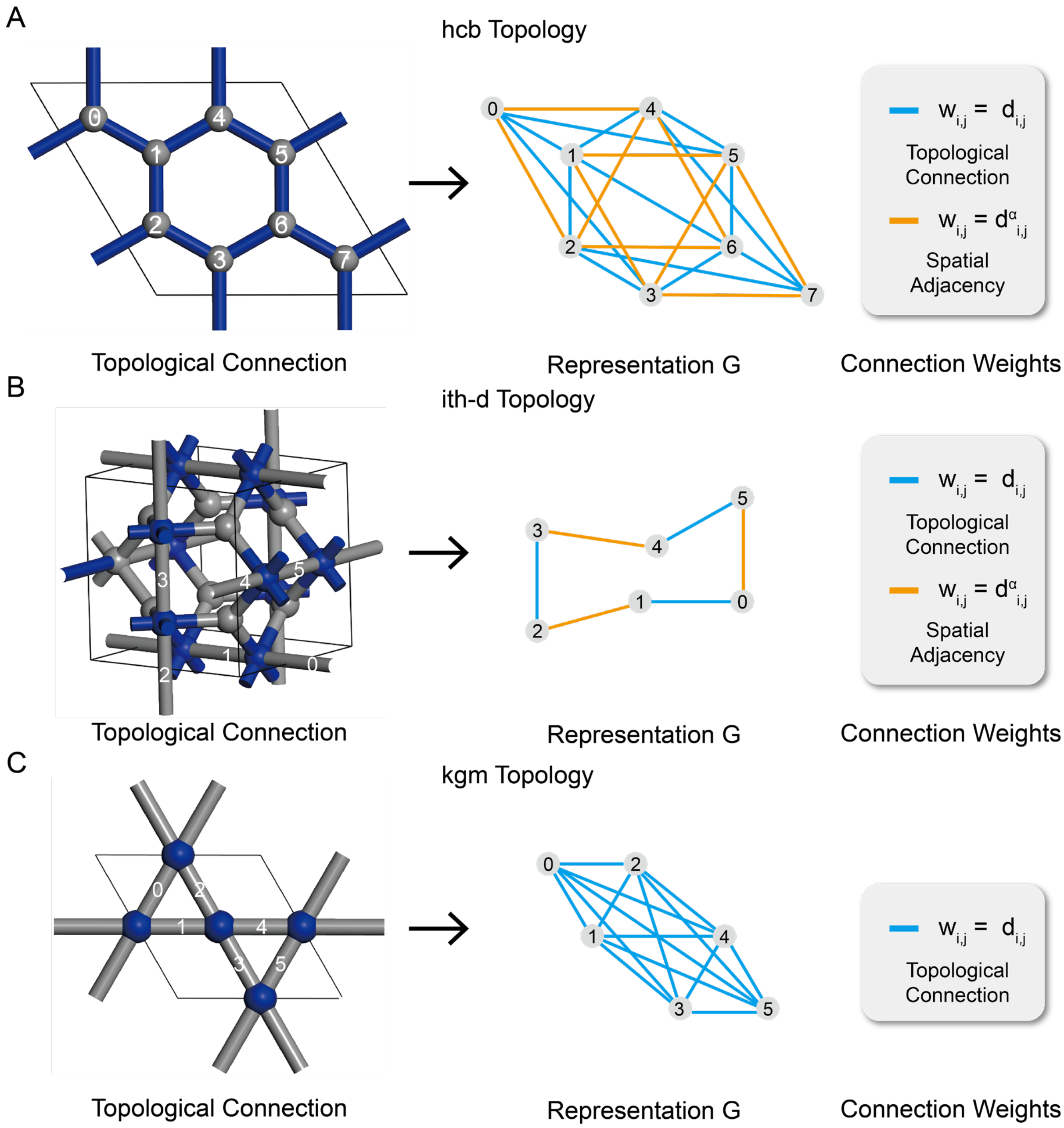

**Fig. S2.** $\boldsymbol{G(i, j, w_{i,j})}$ **of candidate experimental structures for sampling VQE simulation.**

Framework mapping of connections, $(i, j)$, (blue) between building block sites (grey) into a graphical representation ($G(i, j, w_{i,j})$) of **(A)** hcb topology **(B)** ith-d topology **(C)** kgm topology. Each connection is weighted by $w_{i,j}$, which quantifies the strength of either the direct topological connection (light blue) or the spatial adjacency (yellow).

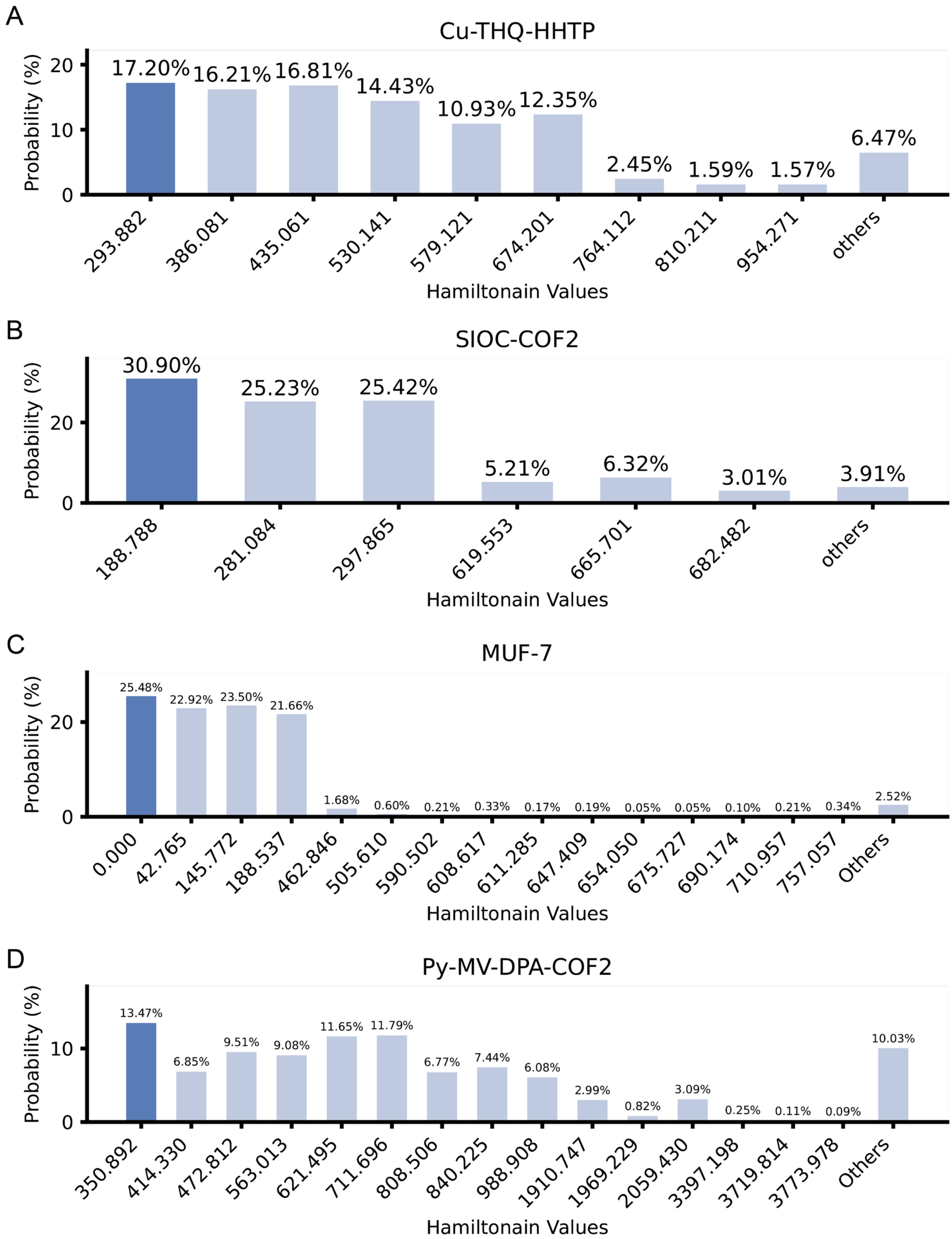

**Fig. S3. Final probability distribution of VQE simulation of candidate structures.** The lowest Hamiltonian of each structure, marked with dark blue, corresponds to the experimental structures. For visual clarity, selective Hamiltonians are shown within individual thresholds.

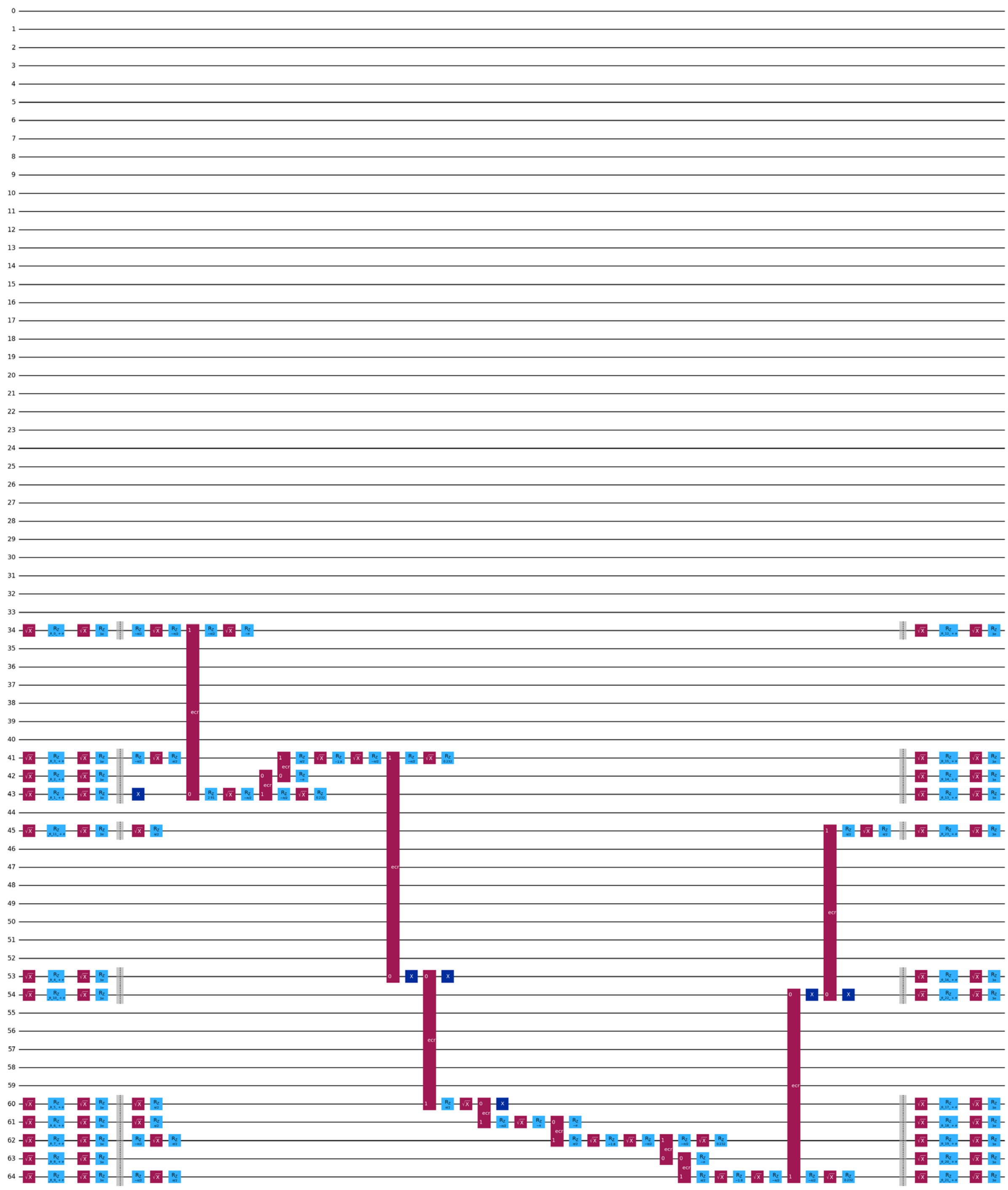

**Fig. S4. Input quantum circuit for the 12-qubit system of SIOC-COF2.** The circuit was transpiled with an optimization level of 3 and executed on ibm_kyiv.

**Note S1. Computational method for Sampling VQE simulation**

To simulate experimental structures, the unit cell of hcb topology was modeled as an eight-linker-site system, and kgm and ith-d topologies were assumed as six-linker-site systems. The characteristic lengths of candidate linkers, $l^t$, were measured by visualizing structures using Atomic Simulation Environment (ASE) (37) (table S2). The spatial distance between nodes, $d_{i,j}$, was determined based on measured distances between nodes from the cgd format of each topology candidate obtained from the RCSR database (38) (table S3). The sensitivity parameter, α, for each experimental structure was determined from the comparative calculations across different α values (table S1).

Once the graph-based framework, $G(i, j, w_{i,j})$, was prepared, the Hamiltonian model was mapped into a quantum circuit using a Two-Local ansatz. The optimization process was executed on IBM Qiskit (26) using the SPSA classical optimizer with a maximum iteration count of 300. Each trial state was sampled 1024 times per iteration to approximate the expectation value of the Hamiltonian. The Minimum Eigen Optimizer was used to solve the optimization problem. The final probability distributions were obtained after 128 iterations for each structure to account for fluctuations in individual runs.

For the Sampling VQE simulation, the total Hamiltonian, $\mathrm{H(q)}$, was formulated with the addition of balancing constants for ratio and occupancy cost terms, $C_{ratio}$ and $C_{occ.}$, as follows:

$$H(q) = C_{ratio} \sum_t \left( \sum_{i=0}^{N_i - 1} {q^t}_i - n_t \right)^2 + C_{occ.} \sum_{i=0}^{N_i - 1} \left( \sum_t {q^t}_i - 1 \right)^2 + \sum_{G \in (i,j,w_{i,j})} w_{i.j} (L(q, G) - \bar{L})^2 \quad (1)$$

The ratio and occupancy cost terms were assigned weighting factors of $C_{ratio} = 200$ and $C_{occ.} = 300$, respectively, to strongly enforce these fundamental structure constraints. These weighting factors were chosen to prevent the balance cost term from dominating the total Hamiltonian, as variations in $w_{i.j}$ and $L(q, G)$ could otherwise disproportionately influence the optimization process. This ensures that the fundamental structural rules for forming a reasonable porous framework are maintained while balancing out the contributions from spatial and connectivity-based constraints.

**Table S1. Results of varying sensitivity parameter, α.** The table shows the number of instances in which the lowest Hamiltonian was observed with the highest probability across 128 individual runs. For example, in the case of MUF-7 at α = 0.5, the lowest Hamiltonian configuration was identified as the most probable (i.e., had the highest probability) 109 times out of 128 runs.

| α | MUF-7 | Cu-THQ-HHTP | Py-MV-DBA-COF |
|---|---|---|---|
| 0.5 | 109 | 85 | 31 |
| 0.25 | 115 | 78 | 28 |
| 0.1 | 116 | 84 | 34 |
| 0.01 | 108 | 86 | 29 |

**Table S2. Characteristic lengths of candidate linkers.**

| MOF | Linker | $l^t$ [Å] |
|---|---|---|
| Cu-THQ-HHTP | THQ | 2.42 |
| | HHTP | 4.87 |
| Py-MV-DBA-COF2 | DBA[12] | 8.027 |
| | DBA[18] | 10.516 |
| MUF-7 | BDC | 2.869 |
| | BPDC | 5.025 |
| SIOC-COF2 | BPDA | 4.6 |
| | TPDA | 6.89 |

**Table S3. Summary of parameters for $G(i, j, w_{i,j})$**

| MOF | Topology | $N_i$ | α | $d_{i,j}$ [Å] | $w_{i,j}$ [Å] |
|---|---|---|---|---|---|
| Cu-THQ-HHTP | hcb | 8 | 1 | 3 | 3 |
| | | | 0.01 | 5.2 | 1.02 |
| Py-MV-DBA-COF2 | hcb | 8 | 1 | 3 | 3 |
| | | | 0.1 | 5.2 | 1.18 |
| MUF-7 | ith-d | 6 | 1 | 3.92 | 3.92 |
| | | | 0.1 | 3.92 | 1.15 |
| SIOC-COF2 | kgm | 6 | 1 | 1.5 | 1.5 |
| | | | | 2.6 | 2.6 |
| | | | | 3 | 3 |